\documentclass[twocolumn,epsbox]{jpsj2}

\newcommand{\simj}{\stackrel{>}{_\sim}}
\newcommand{\simk}{\stackrel{<}{_\sim}}

\def\runtitle{
Superconductivity  and  Spin gap in the zigzag chain $t$-$J$ model  

}
\def\runauthor{Kazuhiro {\sc Sano} and Yoshiaki {\sc \=Ono}}

\title{Superconductivity  and  Spin gap in the zigzag chain $t$-$J$ model   simulating a CuO double chain in  Pr$_{2}$Ba$_{4}$Cu$_{7}$O$_{15-\delta}$
 }

\author{%
Kazuhiro {\sc Sano}\thanks{E-mail address: sano@phen.mie-u.ac.jp} and Yoshiaki {\sc \=Ono}\raisebox{0.5ex}{1,2}
}

\inst{%
Department of Physics Engineering,  Mie University, Tsu, Mie 514-8507       \\ 
\raisebox{0.5ex}{1}Department of Physics, Niigata University, Ikarashi, Niigata 950-2181    \\
\raisebox{0.5ex}{2}Center for Transdisciplinary Research, Niigata University, Ikarashi, Niigata 950-2181
}

\recdate{\today}

\abst{%
Using the numerical diagonalization method, we examine the one-dimensional 
$t_1$-$t_2$-$J_1$-$J_2$ model (zigzag chain $t$-$J$ model) which represents an effective model for metallic CuO double chain in the superconductor Pr$_{2}$Ba$_{4}$Cu$_{7}$O$_{15-\delta}$.  
Based on the Tomonaga-Luttinger liquid theory, we calculate the Luttinger-liquid parameter $K_{\rho}$ as a function of electron density $n$.
It is found  that  superconductivity is realized in  parameter region corresponding   to the experimental result.
We show  phase diagram  of  spin gap  on the $t_2/|t_1|$-$n$ plane  by analyzing the expectation value of  twist-operator $Z_{\sigma}$ in the spin sector. 
The spin gap appears in the region with large $t_2/|t_1|$, where the phase boundary  at half-filling  is consistent  with that of the known frustrated quantum spin system.
The analysis also suggests that the estimated value of the spin gap  reaches $\sim 100K$ in the realistic parameter region of  Pr$_{2}$Ba$_{4}$Cu$_{7}$O$_{15-\delta}$.    
}
\kword{ zigzag chain $t$-$J$ model, spin gap, numerical 
 diagonalization, superconductivity, Tomonaga-Luttinger liquid theory}

\begin{document}

\sloppy
\maketitle


Recently, Matsukawa {\it et al.}  have  discovered a new superconductor Pr$_{2}$Ba$_{4}$Cu$_{7}$O$_{15-\delta}$(Pr247) in which  CuO double chains are considered to derive  the superconductivity at $T_c \sim 20$K\cite{Matsukawa,Yamada}. 
Since  electronic conduction in CuO$_2$ plane  of Pr247 is suppressed due to the so-called Fehrenbacher-Rice state\cite{Rice},   the  double chains are expected to play a  crucial role  for  metallic state of the material.
In fact, anisotropy in the resistivity of a single crystal  shows the one-dimensional(1D)  conductivity  based on the  CuO double chains and the NQR experiment also indicates that the superconductivity is realized in the CuO double chains \cite{Horii,Mizokawa,Watanabe,Sasaki}.
These experiments stimulate  our interest in the  theoretical aspect  for  the electronic state  and  the superconductivity of the double chain system. 

Many theoretical works have been performed on the electronic state of   double chain systems such  as    Ladder models and zigzag chain models.\cite{Balentz,Fabrizio,Emery,Emery1,Solyom,Voit,Doi,Kuroki,Daul,Daul2,Seo,Nishimoto,Ohta,Okunishi,Sano-O-Y,Nakano-Kuro}
Generally speaking, the electronic state of these two-band models is characterized by  existence of four Fermi points, namely, $\pm k_{F_1}$ and $\pm k_{F_2}$ on the Fermi surface. 
In the weak coupling regime,  bosonaization method reveals that  the low-energy excitations  of the double chain   are given by a single gapless charge mode with a gapped spin mode  (labeled as $c1s0$),
 when the ratio of the two Fermi velocities $|v_{F_1}/v_{F_2}|$ is smaller than a critical value $\sim 8.6$.\cite{Balentz,Fabrizio,Emery,c2s2}. 
The correlation functions of the superconductivity(SC)  and that of the charge density wave (CDW)  decay as $\sim r^{-\frac{1}{2K_{\rho}}}$ and $\sim \cos[{2(k_{F_2}-k_{F_1}) r}] r^{-2K_{\rho}}$, respectively, while that of the spin density wave (SDW)  decays exponentially. 
Here, $K_{\rho}$ is the Luttinger liquid parameter and determines the critical exponents of  various types of correlation functions in  the model which is isotropic in spin space.\cite{Emery1,Solyom,Voit} 
In the  $c1s0$ region, the SC correlation is dominant for  $K_\rho >1/2$, while, the CDW correlation is dominant for $K_\rho <1/2$. 

In the strong coupling regime, the double chain systems  have been studied  by using numerical methods.\cite{Doi,Kuroki,Daul,Daul2,Seo,Nishimoto,Ohta,Okunishi}. At half-filling, the system can be described by a Heisenberg model whose ground state is a spin liquid insulator with a finite spin gap\cite{Kuroki,Daul2,Okunishi}. Away from  half-filling, the system becomes a metallic state which maintains a spin gap\cite{Daul,Okunishi}. This behavior is explained by the existence of electron pairs  produced  by the dominant fluctuations of the $4k_F$ charge density wave or the interchain paring fluctuations. 

Among them, there are few works which consider the model just corresponding to the Pr247 except our previous work\cite{Sano-O-Y} and the very recent work using the fluctuation exchange (FLEX) approximation\cite{Nakano-Kuro}.
In our previous paper, we have investigated the superconductivity in the $d$-$p$ double chain model, simulating a CuO double chain of Pr247 where the tight-binding parameters are determined so as to fit the band structure of the local density approximation(LDA). 
On the basis of the Tomonaga-Luttinger liquid theory, we have obtained  $K_{\rho}$ as a function of the electron density $n$. The doping dependence of $K_{\rho}$ is in good agreement with that of $T_c$ in Pr247 \cite{Yamada}  when we assume that   $T_c$ is  a monotonically increasing function of $K_{\rho}$ at $K_{\rho}>1/2$.
However the Hartree-Fock(HF) approximation has been used in this work and  the analysis is limited in the case of the weak coupling region as well as  the FLEX approximation. 

Since the strong correlation effect may play an important role in the electronic state and the superconductivity of Pr247, a nonperturbative and reliable approach is required. In this work, we  employ the numerical diagonalization method for the double chain $t$-$J$ model whose parameters are selected to cover the  realistic  band structure of the CuO double chains.
We calculate  the Luttinger liquid  parameters  $K_{\rho}$  and address the  behavior of the spin gap  in the strong coupling regime  beyond the previous works. 

We consider the following Hamiltonian for the one-dimensional  $t_1$-$t_2$-$J_1$-$J_2$ model(zigzag chain $t$-$J$ model);
\begin{eqnarray} 
H&=&t_1\sum_{i,\sigma}(c_{i,\sigma}^{\dagger} c_{i+1,\sigma}+h.c.) 
+t_2\sum_{i,\sigma}(c_{i,\sigma}^{\dagger} c_{i+2,\sigma}+h.c.) 
 \nonumber \\
&+&J_1\sum_{i,\sigma}({\bf S}_{i} \cdot {\bf S}_{i+1}-\frac{1}{4}n_in_{i+1})
 \nonumber \\
&+&J_2\sum_{i,\sigma}({\bf S}_i \cdot {\bf S}_{i+2}-\frac{1}{4}n_in_{i+2}),
 \nonumber 
\label{tJ-Hamil}  
\end{eqnarray} 
where $c^{\dagger}_{i,\sigma}$ stands for the creation operator of an
 electron with spin $\sigma$ at site $i$  and 
   $n_{i,\sigma}=c_{i,\sigma}^{\dagger}c_{i,\sigma}$.
Here,  $t_1$ is the hopping energy  between the nearest-neighbor  sites and $t_{2}$  is that  between the next nearest-neighbor  sites as shown in Fig.1(a). 
The interaction parameters $J_1$ and $J_2$ stand for   the exchange
  coupling between the nearest-neighbor  sites and between  the next nearest-neighbor  sites, respectively. We take account of the infinite on-site repulsion by removing states with doubly occupied sites from the Hilbert space. 

To determine the hopping energies  $t_1$ and $t_2$, we compare  the noninteracting $t_1$-$t_2$ band  with the $d$-band of the CuO double chain obtained by the LDA band structure in YaBa$_{2}$Cu$_{4}$O$_{8}$(YBCO)\cite{Draxl}.
Here YBCO includes the CuO double chains with the same lattice structure  as those in Pr247. 
As shown in Fig.1(b), both bands are in good agreement  with each other, when we select $t_1=-0.1$eV and $t_2=-0.45$eV\cite{dp-band}. 

As for the   exchange  coupling  energies, $J_1$ is considered to originate in the  exchange  interaction between electrons in the nearest neighbor  $d$-sites.
The value of $J_1$ is  given by the 2nd order perturbation with respect to  the hopping  $t_{dd}$, $i.e.$   $J_1=4t_{dd}^2/U_d$, where $U_d$ is  the on-site Coulomb interaction between $d$ electrons.
When we assume  $t_{dd}=0.12 {\rm eV}$\cite{dp-band} and $U_d=6{\rm eV}$,  $J_1$ is estimated as $0.01{\rm eV}$.
On the other hand, $J_2$ is considered to the superexchange interaction  acting between the next nearest neighbor Cu-sites connecting through O-site.
Our previous study for $d$-$p$ single chain  and Ladder models\cite{effectiveJ}  indicates that the exchange interaction $J/t_{pd}$ is $\sim 0.14$ for $d$-$p$ single chain,   and about 0.09 for  $d$-$p$ Ladder  at $\Delta/t_{pd} \sim 2.6$ and $U_d/t_{pd}=8$.  Here, we  note that  the 4th order perturbation\cite{Matsukawa-Fuku} with respect to $t_{pd}$ overestimates the value of $J/t_{pd}$  in the case of  $\Delta/t_{pd}  \simk 4$.\cite{effectiveJ} 
By reference from the above, we regard the adequate value of $J_2$ as $\sim 0.15 {\rm eV}$.
These values are  close to that of  corresponding  parameters obtained in CuO$_2$ plain system.\cite{Hybertsen}

%
%
\begin{figure}[t]
  \begin{center}
\includegraphics[width=5.8cm]{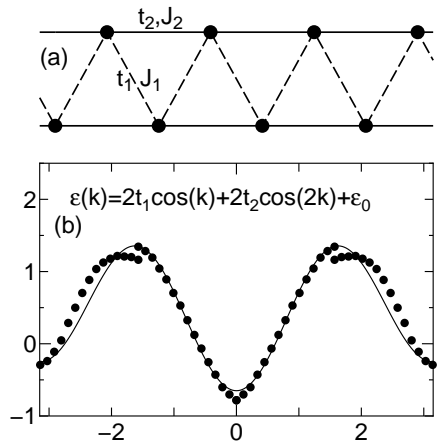}
\end{center}
  \caption[]{
(a) Schematic diagram of  $t_1-t_2-J_1-J_2$  model on  the zigzag chain. 
(b)  Energy dispersion relation for the noninteracting  $t_1-t_2$  model on  the zigzag chain.  Solid lines are the tight-binding result  with $t_1=-0.1 {\rm eV}$ and  $t_2=-0.45 {\rm eV}$. Closed circles are  the LDA result  for the $d$-band of the CuO double chain. 
}
  \label{fig:1}
\end{figure}
%
\begin{figure}[t]
  \begin{center}
\includegraphics[width=5.8cm]{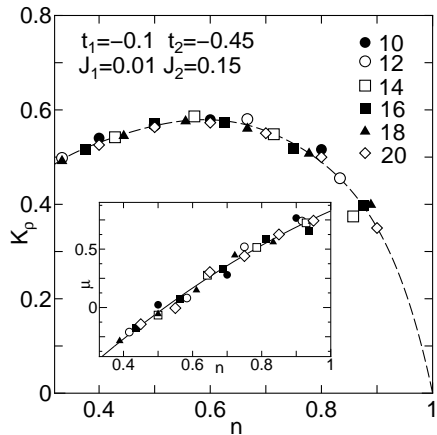}
\end{center}
  \caption[]{
    $K_{\rho}$  as a function of $n$ for  $t_1=-0.1 {\rm eV}$,  $t_2= -0.45 {\rm eV}$,  $J_1=0.01 {\rm eV}$ and $J_2 = 0.15 {\rm eV}$. The  dashed line is a guide for eyes.
Inset shows the chemical potential $\mu$ as a function of $n$. The solid line represents a fitting line  using a  second order polynomial  by the least square method. }
\label{Krow}
\end{figure}
%
%
%

We numerically diagonalize the Hamiltonian  up to 24 sites   using the standard Lanczos algorithm  and  calculate the ground state energy $E_0$.  We use the periodic boundary  condition for $N_e=4m+2$ and the antiperiodic boundary condition for $N_e=4m$, where $N_e$ is the  total number of electrons and $m$ is  an  integer. 
The filling $n$ is defined  by  $n=N_{e}/N$, where $N$ is the total  number of sites.
The critical exponent $K_{\rho}$ is related to  the charge susceptibility
 $\chi_c$ and  the Drude weight $D$ by
$
      K\sb{\rho}=\frac{1}{2}(\pi \chi_c D)^{1/2} ,
$
with 
$
D=\frac{\pi}{N} \frac{\partial^2 E_0(\phi)}{\partial \phi^2}, 
$
where $E_0(\phi)$ is the total energy of the ground state as a function of  
 magnetic flux $N \phi$.\cite{Voit}
Here, the flux is imposed by  introducing the following  gauge transformation:  $c_{m\sigma}{\dagger} \to  e^{im\phi}c_{m\sigma}^{\dagger}$ for an arbitrary site $m$.
 When the charge gap vanishes in the thermodynamic limit, the uniform 
 charge susceptibility $\chi_c$ is obtained from 
$
\chi_c=\frac{\partial n}{\partial \mu},
$
where the chemical potential $\mu (N_e,N)$ is defined by
$
   \mu (N_{e},N)=\frac{E_0(N_e+1,N)-E_0(N_e-1,N)}2.
$
 Using the above $\chi_c$ and $D$, we calculate  the $K_\rho$ from the ground state energy $E_0$ of the finite size system.
%
%

\begin{figure}[t]
  \begin{center}
\includegraphics[width=5.8cm]{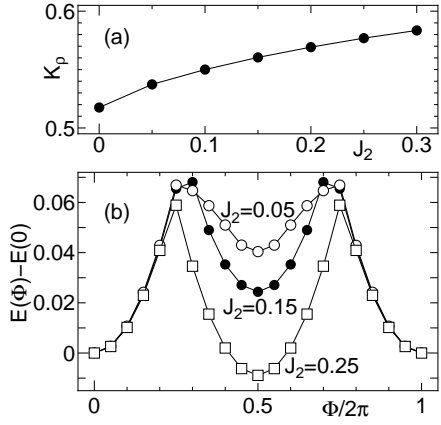}
\end{center}
    \caption[]{
(a) $K_{\rho}$ as a function of $J_2$ for $n=2/3$(12electrons/18sites). 
(b) The energy difference $E_0(\phi)-E_0(0)$  as a function of an external flux $\phi$ for $n=2/3$(12electrons/18sites) at $J_2$=0, 0.15 and 0.30eV with $t_1=-0.1 {\rm eV}$,  $t_2= -0.45 {\rm eV}$ and $J_1=0.01 {\rm eV}$.
  }
\label{fig:3}
\end{figure}
%
Fig. \ref{Krow} shows the Luttinger parameter $K_{\rho}$  as a function of the electron density  $n$ for $t_1=-0.1 {\rm eV}$,  $t_2=-0.45 {\rm eV}$,  $J_1=0.01 {\rm eV}$ and $J_2=0.15 {\rm eV}$. 
Inset shows the chemical potential $\mu$ as a function of $n$, where  data of $\mu$ is fitted  to a second-order polynomial  as a function of $n$ by the least square method and the value of $\chi_c$  is estimated from differential coefficient of the polynomial.
 $K_{\rho}$ increases with increasing $n$ and then have a maximum at an optimal electron density at $n\sim 0.6$.
In the region of $0.4\simk n \simk 0.8$, the value of $K_{\rho}$ exceeds $0.5$ when the SC correlation becomes most dominant as compared with the other correlations (SC phase).
The overall behavior of  $K_{\rho}$  is consistent with our previous work obtained by the HF approximation\cite{Sano-O-Y} except near the half-filling.

Figure \ref{fig:3}(a) shows  the value of $K_{\rho}$ as a function of  $J_2$  for the  12electrons/18sites system.
As $J$ increases, $K_{\rho}$  increases and it  becomes larger than 0.5 even if  $J_2=0$. 
To confirm the superconductivity, we calculate the ground state  energy  $E_0(\phi)$,  as a function of an external flux $\phi$.
As shown in Fig. \ref{Krow}(b),  anomalous flux  quantization  clearly occurs  at $J\simj 0.05$.
It suggests that  the SC phase  appears at $K_{\rho}> 0.5$.

Next, we consider the phase diagram of the spin gap $\Delta_\sigma$.
Generally speaking, it is not easy for  numerical methods to estimate  $\Delta_\sigma$ preciously in the case of the energy scale  being small.  
Especially, it is very difficult to determine the phase boundary of the spin gap which is defined  by $\Delta_\sigma=0$.
To overcome this difficulty, we introduce   $twist$-$operator$  $Z_\sigma$, given as 
\begin{equation}
Z_\sigma=\exp[\frac{2\pi i}{N}\sum_{j=1}^Nj(n_{j\uparrow}-n_{j\downarrow})].
\label{Zs}
\end{equation}
When the expectation value  $\langle Z_\sigma \rangle >0(<0)$,  the system  becomes  spin gapfull(gapless) as  has already  been well examined  by Nakamura $et$ $al$. in the study of 1D extended  Hubbard  model\cite{Nakamura-Voit}. 
We expect that this method is applicable to our zigzag chain model  as well.

\begin{figure}[t]
  \begin{center}
\includegraphics[width=5.8cm]{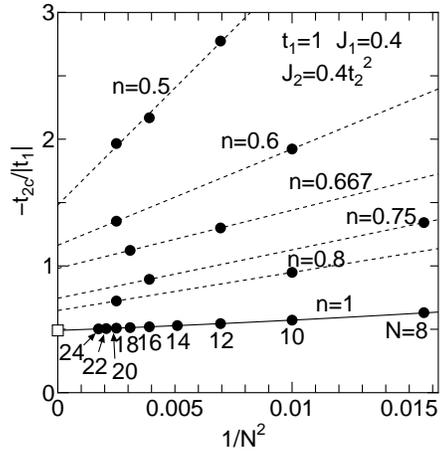}
\end{center}    
\caption[]{
Size dependence of the critical point $t_{2c}$ for $n=1/2,3/5,3/4,4/5$, and 1
  determined by the condition $\langle Z_\sigma \rangle=0$.
The open square stands the well known result of the $J_1$-$J_2$ Heisenberg model in the limit $N \to \infty$. 
}
\label{fig:4}
\end{figure}

In Fig.\ref{fig:4}, we show the size dependence of the critical point $t_{2c}/|t_1|$  determined by  $\langle Z_\sigma \rangle=0$, where we set the relation between  the electron hopping  and the exchange interaction as   $(t_{2}/|t_1|)^2=J_2/J_1$.
 For $n=1$, our system  reduces to the $J_1$-$J_2$ Heisenberg model and the critical point $t_{2c}/|t_1|$ is well   scaled by $1/N^2$.
The extrapolated value $|t_{2c}/t_1|=0.491$ is very close to the  known result of the  Heisenberg model,  $J_{2c}/J_1=(t_{2c}/t_1)^2=0.241$\cite{Okamoto}.
For $n <1$, we obtain the extrapolated value by assuming the same size dependence, where  we set $J_1/|t_1|=0.4$, and $J_2/J_1=(t_2/t_1)^2$ to correspond to 
the zigzag chain Hubbard model($t_1$-$t_2$-$U$ model).

\begin{figure}[!h]
  \begin{center}
\includegraphics[width=5.8cm]{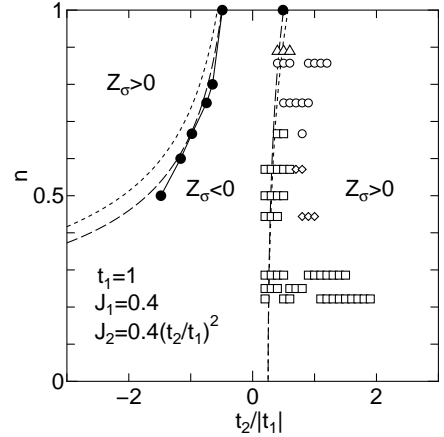}
\end{center}    
\caption[]{
Phase diagram   on  the $t_2/|t_1|$-$n$ plane. The solid circles stand   
the extrapolated value of the critical point $t_{2c}/|t_1|$. 
The open symbols, $\triangle$,  $\bigcirc$, $\diamondsuit$ and $\Box$ indicate  spin polarized  ground state of finite size systems with  $S_{total}/S_{max}=1/4$, $S_{total}/S_{max}=1/3$, $S_{total}/S_{max}=1/2$  and $S_{total}/S_{max}=1$, respectively, where $S_{total}$ is the total spin  of the ferromagnetic state and $S_{max}$ is the maximum possible   value of $S_{total}$.
The broken lines are the phase boundary between c2s2 and c1s0 obtained by the weak coupling theory.\cite{Balentz} 
The dashed lines present the  boundary  between  the region of four  Fermi  points and that of two  Fermi  points in the noninteracting model.  

}
\label{fig:5}
\end{figure}

\begin{figure}[!h]
  \begin{center}
\includegraphics[width=5.8cm]{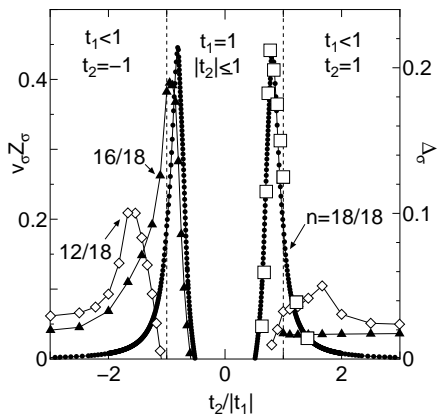}
\end{center}    
\caption[]{
The value of $v_{\sigma}\langle Z_\sigma \rangle$ as a function of $t_2/|t_1|$   at $n=12/18,16/18,18/18$ (18-sites systems with 12,16,18-electrons), where we set $t_2=-1$ for $t_2/|t_1|<-1$, $t_1=1$ for $|t_2/t_1| \le 1$, and $t_2=1$ for $t_2/|t_1|>1$, respectively.
For all values of  $t_2/|t_1|$,  we set $J_1=0.5t_1^2$, and $J_2=0.5t_2^2$ to correspond to the $t_1$-$t_2$-$U$ model with $U=8$.
The open squares stand   the spin gap of $\Delta_{\sigma}$  obtained from the DMRG method for $n=1$\cite{White}.
}
\label{fig:6}
\end{figure}

Fig.\ref{fig:5} shows  the phase diagram  on  the $t_2/|t_1|$-$n$ plane together with the result of the weak coupling theory.
It shows that the phase boundary of the spin gap  is close to that from the weak coupling theory for the region with $n\simj 0.5$ and $t_2/|t_1|\simk -1$.
It also suggests that the boundary is almost independent of $U$ in the  $t_1$-$t_2$-$U$ model, as has already reported for $n=1$.\cite{Otsuka,Torio}
We note that the phase diagram suggests that the parameter region corresponding to Pr247 belongs to the spin gapped phase with $Z_\sigma>0$.

In the region  $ 0.5 \simk t_2/|t_1| \simk 1.9 $, we  find that 
the ground state of the finite system is a spin polarized ferromagnetic state in part as shown in Fig.5\cite{Doi,Daul,Nakano-Taka}, where we use the systems  for $n=4/14,8/14,12/14,4/16,8/16,12/16,4/18,8/18,12/18$ and $16/18$.
The region of the ferromagnetic state is disconnected in respect to $t_2/|t_1|$, which is considered to be caused by a finite size effect. 
Although the phase boundary of $\langle Z_\sigma \rangle=0$  is masked by the ferromagnetic state in the region, we have confirmed that the sign of $\langle Z_\sigma \rangle$ is positive for $t_2/|t_1|\simj 1$ and  negative for $t_2/|t_1|\simk 0$ except the ferromagnetic phase.

Finally, we consider the relationship between   $\langle Z_\sigma \rangle$  and  the value of the spin gap $\Delta_{\sigma}$.
It is known that $\langle Z_\sigma \rangle$  corresponds to the expectation  value of the nonlinear term $\cos(\surd{8\phi_{\sigma}})$ in the sine-Gordon model which is  the effective Hamiltonian of 1D electron system\cite{Nakamura-Voit}.
Because this term becomes a source producing the gap,  there is a close relation  between  $\Delta_{\sigma}$ and  $\langle Z_\sigma \rangle$\cite{Lukyanov}.
We find  that $\Delta_{\sigma}$ of the infinite system is  almost proportional to the product of $ v_{\sigma}$ and $\langle Z_\sigma \rangle$ of the 18-sites system in  wide range of the parameter  $t_2/|t_1|$ at $n=1$,
 where $v_{\sigma}$ is the spin  velocity corresponding to the energy scale of the spin part of the  effective Hamiltonian\cite{vs}. 
In Fig.\ref{fig:6}, we plot $\langle Z_\sigma \rangle$ in  the 18-sites  system together with $\Delta_{\sigma}$  obtained by DMRG method for $n=1$.\cite{White}

Remarkably, a phenomenological relation, $\Delta_{\sigma}=0.48v_{\sigma}\langle Z_\sigma \rangle$, is observed at $n=1$ for all values of  $t_2/|t_1|$.
Assuming the same relation is  satisfied even for  $n<1$, 
we estimate the spin  gap $\Delta_{\sigma}$ from the value of  $v_{\sigma}\langle Z_\sigma \rangle$ of  the 18-sites  system.
To confirm this  assumption, we compare our result with  $\Delta_{\sigma}$ of the $t_1$-$t_2$-$U$ model obtained by the recent DMRG method.\cite{Okunishi} 
As shown in Fig.6, we obtain  $\langle Z_\sigma \rangle \sim 0.25$ and $v_s \sim 1.5$ resulting in  $\Delta_{\sigma} \sim 0.18$ for $n=16/18$ (18-sites system with 16-electrons) with the  parameters: $|t_1|=1.0$, $t_2=-1.0$,  $J_1=0.5$, and $J_2=0.5$.
The result is in good agreement with  the  DMRG result, i.e., $\Delta_{\sigma} \sim 0.16$ of  the corresponding $t_1$-$t_2$-$U$ model with  $|t_1|=1.0$, $t_2=-1.0$ and  $U=8$.\cite{Okunishi}
Then, we expect that our  analysis is  useful to estimate   the spin gap, even for $n<1$.
When we apply the above method to the realistic parameter region of Pr247, we obtain   $\Delta \simeq 0.0072$eV for $n=16/18$ with  $t_1=-0.1$eV, $t_2=-0.45$eV,  $J_1=0.01$eV and $J_2=0.15$eV.
In the case of  $n=12/18$, we  find  $\Delta_{\sigma} \simeq 0.011$eV. 
These results suggest that the order of the spin gap amounts to  $\Delta_{\sigma}\sim 100$K  in the realistic parameter region of Pr247  and  is  larger than $T_c \simeq 20$K.


In summary, we investigate the one-dimensional $t_1$-$t_2$-$J_1$-$J_2$ model as an effective model for metallic CuO double chain of Pr247 using the numerical diagonalization method.
The  hopping  parameters of electron  are chosen so as to fit the $d$-band from the CuO double chain obtained from LDA calculation and the exchange  coupling  energies $J_1$ and $J_2$ are estimated by  the known results of 1D $d$-$p$ models.
In a realistic parameter region, we show that the Luttinger liquid parameter $K_{\rho}$  is greater than 0.5 and  the anomalous flux  quantization is found.  
It suggests that the CuO double chain is responsible for the superconductivity of Pr 247 on the basis of the Tomonaga-Luttinger liquid theory.
We also calculate the expectation value of the twist-operator $Z_{\sigma}$ and
obtain the phase boundary of the spin gap  including the region with large value of $t_2/|t_1|$.
By comparing with  the known result of the $J_1$-$J_2$ Heisenberg model,
we  estimate   the value of the spin gap $\Delta_{\sigma}$ through $\langle Z_{\sigma} \rangle$ and spin velocity $v_{\sigma}$, and find the spin gap becomes  $\sim 100K$ in the realistic parameter region of Pr247. 
This result is consistent with the recent NQR  experiment, where $(T_1T)^{-1}$ is suppressed in the superconducting sample as compared with the non-superconducting sample even above $T_c$.

\acknowledgements
The authors thank Y. Yamada, K. Okunishi, K. Kuroki and T. Nakano for useful discussion.
This work is partially supported by the Grant-in-Aid for  Scientific Research from the Ministry of Education, Culture, Sports, Science and Technology of Japan.



\begin{thebibliography}{99}
\bibitem{Matsukawa}
M. Matsukawa, Yuh Yamada, M. Chiba, H. Ogasawara, T. Shibata, A. Matsushita and Y. Takano: Physica C {\bf 411} (2004) 101.
%
%
\bibitem{Yamada}
Yuh Yamada and A. Matsushita:  Physica C{\bf 426-431} (2005) 213.
%
\bibitem{Rice}
R. Fehrenbacher and T. M. Rice: Phys. Rev. Lett. {\bf 70} (1993) 3471.
%
\bibitem{Horii}
S. Horii, U. Mizutani, H. Ikuta, Yuh Yamada, J. H. Ye, A. Matsushita, N. E. Hussey, H. Takagi and I. Hirabayashi: Phys. Rev. B {\bf 61} (2000) 6327. 
%
\bibitem{Mizokawa}
T. Mizokawa, C. Kim, Z-X. Shen, A. Ino, T. Yoshida, A. Fujimori, M. Goto, H. Eisaki, S. Uchida, M. Tagami, K. Yoshida, A. I. Rykov, Y. Siohara, K. Tomimoto, S. Tajima, Yuh Yamada, S. Horii, N. Yamada, Yasuji Yamada, I. Hirabayashi: Phys. Rev. Lett. {\bf 85} (2000) 4779.
%
\bibitem{Watanabe}
S. Watanabe, Yuh. Yamada and S. Sasaki:   Physica C {\bf 426-431} (2005) 473. 
%
\bibitem{Sasaki}
S. Sasaki, S. Watanabe, Y. Yamada, F. Ishikawa and S. Sekiya: cond-mat/0603067.
%
\bibitem{Balentz} 
L. Balentz and M.P.A. Fisher: Phys. Rev. B {\bf 53}  (1996) 12133.
%
\bibitem{Fabrizio}
M. Fabrizio, Phys. Rev: B {\bf 54} (1996) 10054.
%
\bibitem{Emery}
V. J. Emery, S. A. Kivelson and O. Zachar: Phys. Rev. B {\bf 59}  (1999) 15641.
%
%
\bibitem{Emery1}
  V. J. Emery, in {\it Highly Conducting One-Dimensional Solids},
 edited by J. T. Devreese, R. Evrand and V. van Doren, (Plenum, New
	York, 1979), p.327.
%
%
\bibitem{Solyom}
 J. S\'olyom, Adv. Phys. {\bf 28}, 201 (1979).
%
\bibitem{Voit}
J. Voit, Rep. Prog. Phys. {\bf 58},  977 (1995).
%
%
\bibitem{Doi}
 I. Doi, K. Sano and K. Takano: Phys. Rev. {\bf B45} (1992) 274.
%
%
\bibitem{Kuroki}
K. Kuroki, R. Arita and H. Aoki: J. Phys. Soc. Jpn. {\bf 66} (1997) 3371.
%
\bibitem{Daul} S. Daul and R. M. Noack: Phys. Rev. {\bf B58} (1998) 2635. 
%
\bibitem{Daul2} S. Daul and R. M. Noack, Phys. Rev. {\bf B61} (2000) 1646. 
%
%
\bibitem{Seo}
H. Seo and M. Ogata: Phys. Rev. B {\bf 64} (2001) 113103.
%
%
\bibitem{Nishimoto}
S. Nishimoto  and Y. Ohta: Phys. Rev. B {\bf 68} (2003) 235114.
%
\bibitem{Ohta}
Y. Ohta, S. Nishimoto, T. Shirakawa, Y. Yamaguchi: Phys. Rev. B {\bf 72} (2005) 012503.
%
\bibitem{Okunishi}
K. Okunishi: Phys. Rev. {\bf B 75} (2007) 174514.
%
%
%
\bibitem{Sano-O-Y}
K. Sano, Y. \=Ono, Yuh Yamada: J. Phys. Soc. Jpn. 74, (2005) 2885.
%
\bibitem{Nakano-Kuro}
  T. Nakano, K. Kuroki, S. Onari:   cond-mat/0701160. 
%
\bibitem{c2s2}
When  the ratio  $|v_{F_1}/v_{F_2}|$  is larger  than the critical value $\simeq 8.6$, the low energy excitations are given by two gapless charge modes and two gapless spin modes (labeled as $c2s2$). 
%
\bibitem{Draxl}
C. Ambrosch-Draxl, P. Blaha and K. Schwarz: Phys. Rev. B {\bf 44} (1991) 5141.
%

\bibitem{dp-band}
The realistic tight-binding parameters of the $d$-$p$  double chain  band have been  estimated as $t_{pd}=1.6 {\rm eV}$,  $t_{pp}= 0.43 {\rm eV}$,  $t_{dd}=  0.12 {\rm eV}$ and $\Delta = 4.1 {\rm eV}$, respectively\cite{Sano-O-Y}. Here,  $t_{pd}$ is the hopping energy  between the nearest-neighbor $d$,  $p$ sites and $t_{pp}$ ($t_{dd}$) is the hopping  energy  between the nearest-neighbor $p$ ($d$) sites and  $\Delta$ is the charge-transfer energy, respectively.
%
%
\bibitem{effectiveJ}
K. Sano, Y. \=Ono, J. Phys. Soc. Jpn. 71 (2002) Suppl. 353.
%
%
\bibitem{Matsukawa-Fuku}
 H. Matsukawa  and H. Fukuyama: 
J. Phys. Soc. Jpn. {\bf 58}  (1989) 2845. 
%
\bibitem{Hybertsen}
 M. S. Hybertsen, E. B. Stechel, M. Schluter and D. R. Jennison,  Phys. Rev. {\bf B41} (1990) 11068. 
%
\bibitem{Nakamura-Voit} M. Nakamura and J. Voit,  Phys. Rev. {\bf B65} (2002) 153110. 
%
\bibitem{Okamoto} K. Okamoto and K. Nomura, Phys. Lett. A {\bf 169} (1992) 433.
%
\bibitem{Otsuka} H. Otsuka  Phys. Rev. B {\bf 57}  (1998) 14658.
%
\bibitem{Torio} M. E. Torio, A. A. Aligia and H. A. Ceccatto, Phys. Rev. {\bf B67} (2003) 165102.
%
\bibitem{Nakano-Taka} H. Nakano and Y. Takahashi: J. Phys. Soc. Jpn. {\bf 72} (2003) 1191.
%
%
\bibitem{vs}
The value of $v_{\sigma}$ is calculated by the energy difference between the first excited triplet state with wave number $k=2\pi/N$ and the ground state of  finite systems.
%
\bibitem{sine-Golrdon}
In the low energy limit, the effective Hamiltonian of the spin part  is 
 given by
$     H_{\sigma}=\frac{v_{\sigma}}{2\pi}\int_0^L {\rm d}x
  \left[K_{\sigma}(\partial_x \theta_{\sigma})^2
       +K_{\sigma}^{-1}(\partial_x \phi_{\sigma})^2\right]  + \frac{2 g_{3\perp}}{(2\pi\alpha)^2}
  \int_0^L {\rm d}x \cos[\sqrt{8}\phi_{\sigma}(x)], $
where  $K_{\sigma}$ and $g_{3\perp}$  are  the coupling  parameter
and backward scattering term, respectively. 
%
%
\bibitem{Lukyanov}
S. Lukyanov and A. Zamolodchikov:  Nuclear Physics B493 (1997) 571.
%
\bibitem{White}
S. R. White  and I. Affleck: Phys. Rev. B {\bf 54} (1996) 9862.
%


%
\end{thebibliography}
\end{document}